\documentclass[aps,showpacs,10pt]{revtex4}

\usepackage{graphicx}
\usepackage{amsmath}
\usepackage{amssymb}

\newcommand{\be}{\begin{equation}}
\newcommand{\ee}{\end{equation}}
\newcommand{\ba}{\begin{eqnarray}}
\newcommand{\ea}{\end{eqnarray}}

\newcount\bozza \bozza=0
\ifnum\bozza=1
\newdimen\shift \shift=-2truecm
\def\lb#1{%
{\label{#1}\rlap{\kern\shift{$\scriptstyle#1$}}}}
\else\def\lb#1{\label{#1}} \fi

\begin{document}


\title{Effects of energy dependence in the quasiparticle density
of states on far-infrared absorption in the pseudogap state}

\author{S.G.~Sharapov}
\email{sharapo@mcmaster.ca}
\author{J.P.~Carbotte}
\email{carbotte@mcmaster.ca}

\affiliation{Department of Physics and Astronomy, McMaster University,
        Hamilton, Ontario, Canada, L8S 4M1}

\date{\today }

\begin{abstract}
We derive a relationship between the optical conductivity scattering
rate $1/\tau(\omega)$ and the electron-boson spectral function
$\alpha^2F(\Omega)$ valid for the case when the electronic density
of states, $N(\epsilon)$, cannot be taken as constant in the
vicinity of the Fermi level. This relationship turns out to be
useful for analyzing the experimental data in the pseudogap state of
cuprate superconductors.
\end{abstract}

\pacs{74.72.-h,78.20.Bh,72.10.Di}



\maketitle

\section{Introduction}

Optical conductivity data on $\sigma(\omega)$ vs $\omega$ contains
important information on electron dynamics. In general however the
relationship between $\sigma(\omega)$ and the electron self-energy
is rather complicated. Under such circumstances the reduction, if
possible, of the complete expression for conductivity, to somewhat
more approximate but simpler analytic form can be valuable. It can
help our understanding of the basic physics as well as provide
experimentalists with a simpler basis for the analysis of data.

In the specific case of an electron-phonon system, the analytic
formula provided by Allen \cite{Allen1971PRB} has proved to be
valuable. It relates the measured optical scattering rate through a
simple integral to the underlying electron-phonon spectral function
$\alpha^2 F(\omega)$ which is, in the end,  the fundamental quantity
of interest. A generalization of Allen's formula to finite
temperature was provided by Shulga {\em et. al.}
\cite{Shulga1991PC}. In this paper we want to extend this previous
work to the case when the underlying electronic density of states is
energy dependent rather than constant. A motivation for this
extension is to provide guidance in the interpretation and analysis
of optical data in the pseudogap regime of the cuprates. In fact the
formula derived herein has already been used in the experimental
work of Hwang {\em et. al.} \cite{Hwang2005} on ortho-II YBCO.

\section{Preliminaries}

The Drude formula for optical conductivity $\sigma(\omega) =
\sigma_1(\omega)+ i \sigma_2(\omega)$ \be \label{sigma-Drude}
\sigma(\omega,T) = \frac{\omega_p^2}{4 \pi} \frac{1}{1/\tau - i
\omega} \ee can be extended (see
\cite{Timusk2003SSC,Puchkov1996JPCM} and Refs. therein) to include a
frequency dependent scattering rate
\begin{equation}
\label{sigma-generalized} \sigma(\omega,T) = \frac{\omega_p^2}{4
\pi} \frac{1}{1/\tau(\omega,T) - i \omega[1+ \lambda(\omega,T)]},
\end{equation}
where $1/\tau(\omega,T)$ is the frequency-dependent optical
scattering rate and $\lambda(\omega,T)$ is the optical mass
enhancement. For a spherical Fermi surface the plasma frequency
$\omega_p^2 =4 \pi n e^2/m$, where $n$ is the free-carrier density
and $m$ is the carrier  mass.

One can solve Eq.~(\ref{sigma-generalized}) for $1/\tau(\omega)$ and
$1+\lambda(\omega)$ in terms of the optical conductivity found from
experiment, \begin{equation} \label{tau_via_sigma}
\frac{1}{\tau(\omega)} = \frac{\omega_p^2}{4 \pi} \mbox{Re} \left(
\frac{1}{\sigma(\omega)}\right)
\end{equation} and
\begin{equation}
\label{lambda_via_sigma} 1+ \lambda(\omega) = -\frac{\omega_p^2}{4
\pi} \frac{1}{\omega}\mbox{Im} \left(
\frac{1}{\sigma(\omega)}\right).
\end{equation}
The plasma frequency can be extracted from the experimental data
using the sum rule $ \int_0^\infty \sigma_1(\omega) d \omega =
\omega_p^2/8. $ Although the representations of experimental data
using $\sigma_1(\omega)$, $\sigma_2(\omega)$ and $1/\tau(\omega)$
with $1+\lambda(\omega)$ are formally equivalent, it has become
rather popular to discuss the pseudogap behavior in high-temperature
superconductors (HTSC) using the language of the optical scattering
rate and mass enhancement. For example, a drop of $1/\tau(\omega)$
extracted from in-plane optical conductivity measurements in HTSC
which is observed below a certain frequency for the temperatures $T<
T_{ab}^{\ast}$ is associated with the above-mentioned pseudogap
\cite{Timusk2003SSC}.

Another advantage of $1/\tau(\omega)$ is that in an electron-phonon
system it is related to the electron-phonon interaction spectral
density, $\alpha^2 F(\omega)$. For example, there is an approximate
relationship \cite{Carbotte.book}
\be \label{alphaF-Frank} \alpha^2 F(\omega) =  \frac{1}{2 \pi}
\frac{d^2}{d \omega^2} \left(\omega \frac{1}{\tau(\omega)} \right)
\ee
which is valid at $T=0$ in the normal state. Note that one can
consider a general electron-boson interaction function, because
instead of $\alpha^2 F(\omega)$ one can, for instance, introduce
electron-spin excitation spectral density, $I^2 \chi(\omega)$
\cite{Carbotte1999Nature}, Thus below we imply this more general
case, but preserve the historical notation $\alpha^2 F(\omega)$.

Despite its ``magic'' simplicity,  formula (\ref{alphaF-Frank})
works rather well (see e.g.
\cite{Marsiglio1998PLA,Marsiglio1999JSC}), but unfortunately its
practical applications are limited because the experimental data
must be (ambiguously) smoothed ``by hand'' before the second
derivative can be taken.   This problem was very recently
circumvented in Ref.~\cite{Dordevic2005PRB}  by solving the
corresponding integral equations for $\alpha^2 F(\omega)$. For
example,  instead of differentiating $\omega/\tau(\omega)$, one can
extract $\alpha^2 F(\omega)$ using the well-known result of Allen
\cite{Allen1971PRB}
\begin{equation}
\label{tau-Allen}
\frac{1}{\tau(\omega)} = \frac{2\pi}{\omega}
\int_{0}^{\omega} d \Omega (\omega-\Omega) \alpha^2 F (\Omega).
\end{equation}
The last expression was
derived using the second order perturbation theory and in fact,
Eqs.~(\ref{tau-Allen}) and (\ref{alphaF-Frank})
are equivalent under the condition $1/\tau(\omega=0) =0$.

A finite temperature generalization of
Eq.~(\ref{tau-Allen}) was derived in Ref.~\cite{Shulga1991PC}
using the Kubo formula
\begin{equation}
\label{tau-Shulga} \frac{1}{\tau(\omega)} = \frac{\pi}{\omega}
\int_{0}^\infty d \Omega \alpha^2 F(\Omega) \left[2 \omega \coth
\frac{\Omega}{2T} - (\omega+\Omega) \coth \frac{\omega+\Omega}{2T}
+ (\omega-\Omega) \coth \frac{\omega-\Omega}{2T} \right].
\end{equation}
In fact, there is no  finite temperature equivalent of a``magic''
equation (\ref{alphaF-Frank}) for Eq.~(\ref{tau-Shulga}), so that
for finite $T$'s $\alpha^2 F$ can only be found by inversion of the
last equation. The numerical method of inversion of
Eq.~(\ref{tau-Shulga}), its limitations and the resulting doping and
temperature dependences of the bosonic spectral function, $\alpha^2
F(\omega)$ in several families of HTSC were investigated in detail
in Ref.~\cite{Dordevic2005PRB}.

There is, however, an important assumption used in deriving both the
$T=0$ equation (\ref{tau-Allen}) and its finite temperature
extension (\ref{tau-Shulga}), viz. the {\em electronic density of
states (DOS), $N(\epsilon)$ is taken as a constant in the vicinity
of the Fermi level.\/} As shown in Ref.~\cite{Dordevic2005PRB} when
Eq.~(\ref{tau-Shulga}) is used for analyzing the experimental data
in the pseudogap state of HTSC, the resulting $\alpha^2 F$ contains
nonphysical negative values. This problem originates from the fact
that the above-mentioned assumption $N(\epsilon) = \mbox{const}$ is
definitely strongly violated in the pseudogap state. The essence of
the pseudogap phenomenon is that $N(\epsilon)$ becomes a nontrivial
function of energy and the form of $N(\epsilon)$ depends strongly on
the temperature and doping.

Therefore it would be very useful to have a generalization
of Eqs.~(\ref{tau-Allen}) and (\ref{tau-Shulga}) valid
for $N(\epsilon) \ne \mbox{const}$. Interestingly, such
a generalization of the zero temperature expression
(\ref{tau-Allen}) was already done
twenty years ago by Mitrovi{\'c} and Fiorucci \cite{Mitrovic1985PRB}
in relation to A15 compounds
\begin{equation}
\label{tau-Mitrovic}
\frac{1}{\tau(\omega)} = \frac{2\pi}{\omega} \int_{0}^{\omega}
d \Omega \alpha^2 F(\Omega) \int_{0}^{\omega-\Omega} d \epsilon
\frac{1}{2}\left[\frac{N(\epsilon)}{N(0)} + \frac{N(-\epsilon)}{N(0)} \right].
\end{equation}
Eq.~(\ref{tau-Mitrovic}) was derived using the method of
Ref.~\cite{Allen1971PRB} and it is easy to see that for
$N(\epsilon) =\mbox{const}$ it reduces to Eq.~(\ref{tau-Allen}).

The purpose of the present work is to obtain a generalization of the
finite temperature expression (\ref{tau-Shulga}) valid for
$N(\epsilon) \ne \mbox{const}$. In contrast to
Refs.~\cite{Allen1971PRB,Mitrovic1985PRB}, we base our derivation on
the Kubo formula which turns out to be more useful for considering
the $T \ne 0$ case.

We begin by presenting in Sec.~\ref{sec:conductivity} the expression
for optical conductivity $\sigma(\omega)$ in terms of frequency
dependent self-energy $\Sigma(\omega)$ for the case $N(\epsilon) \ne
\mbox{const}$. The corresponding expression for $\Sigma(\omega)$ is
obtained in Sec.~\ref{sec:self-energy} for the case of non-constant
{\em quasiparticle DOS} $\tilde{N}(\epsilon)$. The difference
between the usual DOS $N(\epsilon)$ and the quasiparticle DOS
$\tilde{N}(\epsilon)$ is pointed out. In Sec.~\ref{sec:tau-new} we
present the relationship between the optical scattering rate,
$1/\tau(\omega)$, electron-boson interaction function $\alpha^2
F(\omega)$ and the quasiparticle DOS $\tilde{N}(\epsilon)$. The
frequency dependent impurity scattering rate
$1/\tau_{\mathrm{imp}}(\omega)$ is considered in
Sec.~\ref{sec:tau-imp}. In the Discussion,
Sec.~\ref{sec:discussion}, we illustrate that the pseudogap opening
results in the decrease of $1/\tau(\omega)$ and consider possible
applications of our results.

\section{Optical conductivity for $N(\xi)\ne \mbox{const}$}
\label{sec:conductivity}

We begin with the Kubo formula for electrical conductivity
\cite{Yanagisawa.book}
\begin{equation}
\label{sigma-total}
\sigma_{ij}(\omega) = \frac{i}{\omega+i0} \left[ \tau_{ij}- \Pi_{ij}^R(\omega+i0)\right],
\end{equation}
where $\tau_{ij}$ is the diamagnetic term and $\Pi_{ij}(\omega)$ is
the retarded correlation function obtained by analytical
continuation [$\Pi_{ij}^R(\omega) = \Pi_{ij}(i\Omega_m \to
\omega+i0)$] of the imaginary time expression
\begin{equation}
\Pi_{ij} (i \Omega_m) = \frac{1}{V}\int_0^\beta d \tau e^{i \Omega_m \tau}
\langle j_i(\mathbf{q}=0, \tau) j_j(\mathbf{0},0)\rangle,
\qquad \Omega_m = \frac{2 \pi m}{\beta}.
\end{equation}
Here $j_i(\mathbf{q},\tau)$ is the Fourier transform of the
paramagnetic electric current density operator, $V$ is the volume of
the system, $T = 1/\beta$ and for the case of parabolic band
$\tau_{ij}= \omega_p^2/4 \pi \delta_{ij}= e^2 n\delta_{ij}/m$. The
electrical conductivity (\ref{sigma-total}) consists of a regular
part
\begin{equation}
\label{sigma-reg}
\sigma_{ij}^{\mathrm{reg}}(\omega) =
-\frac{i}{\omega}\left[\Pi_{ij}^{R}(\omega)-\Pi_{ij}^R(0) \right]
\end{equation}
and singular part, related to the superconducting condensate,
\begin{equation}
\label{sigma-sin} \sigma_{ij}^{\mathrm{SC}} (\omega) =
\frac{i}{\omega+i0}[\tau_{ij} - \Pi_{ij}^R(0)].
\end{equation}
Since in what follows we restrict ourselves to considering the
normal state of the isotropic two dimensional system, we suppress
the subscript $\mathrm{reg}$ and consider $\sigma(\omega) =
\sigma_{xx}^{\mathrm{reg}}(\omega) =
\sigma_{yy}^{\mathrm{reg}}(\omega)$ deriving, for example,
$\Pi_{xx}(\omega)$.

Neglecting vertex corrections, the calculation of $\sigma(\omega)$
reduces to evaluation of the bubble diagram
\begin{equation}
\label{bubble}
\Pi_{ij} (i \Omega_m) =- 2 e^2 T \sum_{n=-\infty}^{\infty} \int
\frac{d^2 k}{(2 \pi)^2}  v_{Fi}(\mathbf{k}) v_{Fj}(\mathbf{k})
G(i\omega_n+i\Omega_m,\mathbf{k}) G(i\omega_n,\mathbf{k}),
\end{equation}
where $v_{Fi}(\mathbf{k}) = \partial \xi(\mathbf{k})/\partial k_i$
is the Fermi velocity and
\begin{equation}
\label{Green}
G(i \omega_n,\mathbf{k}) =
\frac{1}{i\omega_n - \xi(\mathbf{k})-\Sigma(i\omega_n,\mathbf{k})}
\end{equation}
is the fermionic Green's function with the self-energy
$\Sigma(i\omega_n,\mathbf{k})$ and $\omega_n = \pi(2n+1)/\beta$.
Using the spectral representation
for the Green's function (\ref{Green}), one can easily sum over
fermionic Matsubara frequencies in Eq.~(\ref{bubble})
\begin{equation}
\label{bubble-spectral}
\Pi_{xx}(i \Omega_m) =- 2  e^2 \int_{-\infty}^{\infty} d \xi
\int \frac{d^2k}{(2 \pi)^2}
\delta(\xi - \xi(\mathbf{k}))
 v_{Fx}^2(\mathbf{k})
\int_{-\infty}^{\infty} d \omega_1  \int_{-\infty}^{\infty} d \omega_2
\frac{n_F(\omega_1) -n_F(\omega_2)}{\omega_1-\omega_2+i\Omega_n}
A(\omega_1,\mathbf{k}) A(\omega_2,\mathbf{k}),
\end{equation}
where the spectral function
\begin{equation}
\label{SF} A(\omega,\mathbf{k}) = -\frac{1}{\pi} \mbox{Im}
G_R(\omega+i0,\mathbf{k}),
\end{equation}
$n_F(\omega)=1/(e^{\beta \omega}+1)$ is the Fermi distribution, and
we inserted the integral over $\xi$ which is equal to 1. Since we
are interested in the complex conductivity $\sigma(\omega)$ we
cannot remove one of the integrations over $\omega$ in
(\ref{bubble-spectral}) by taking $\mbox{Im} \Pi(\omega+i0)$, but
one of the integrations can be done by again using the spectral
representation for retarded (advanced) Green's function
\begin{equation}
G_{R,A}(\omega,\mathbf{k}) = \int_{-\infty}^{\infty}d \omega_2
\frac{A(\omega_2,\mathbf{k})}{\omega\pm i0-\omega_2}.
\end{equation}
Then we obtain
\begin{equation}
\label{bubble-vE} \Pi_{xx}(\omega) = -  2  e^2
\int_{-\infty}^{\infty} d \omega^\prime
n_F(\omega^\prime)\int_{-\infty}^{\infty} d \xi N(\xi) v_{Fx}^2
(\xi) A(\omega^\prime,\xi) [G_R(\omega^\prime+\omega,\xi)
+G_A(\omega^\prime-\omega,\xi)],
\end{equation}
where to isolate the effects of the energy dependence of the
single-spin-band DOS,
\begin{equation}
\label{DOS-band} N(\xi) = \int \frac{d^2 k}{(2\pi)^2}
\delta(\xi-\xi(\mathbf{k})),
\end{equation}
the velocity $v_{Fx}^2(\xi)$ is defined as (see
Refs.~\cite{Pickett1980PRB,Mitrovic1985PRB})
\begin{equation}
v_{Fx}^2(\xi) \equiv \frac{1}{N(\xi)}\int \frac{d^2 k}{(2\pi)^2}
v_{Fx}^2(\mathbf{k}) \delta(\xi-\xi(\mathbf{k})).
\end{equation}
Writing Eq.~(\ref{bubble-vE}) we also assumed that
$A(\omega,\mathbf{k})$ and, accordingly,
$G_{R,A}(\omega,\mathbf{k})$ to be dependent only on
$\xi(\mathbf{k})$.

Now we must make two important assumptions: The first one is quite
common and states that the self-energy $\Sigma(i\omega,\xi)$ does
not depend on $\xi$, so that the whole dependence of
$G_{R,A}(\omega,\xi)$ is contained in the free-electron dispersion
$\xi(\mathbf{k})$. The second assumption is that the energy
dependence of the square of the plasma frequency
\begin{equation}
\label{plasma-xi} \frac{\omega_p^2(\xi)}{4 \pi} = 2e^2 N(\xi)
v_{Fx}^2(\xi)
\end{equation}
can be ignored as compared to the dependence of $N(\xi)$ in the
vicinity of $\xi=0$, so that in Eq.~(\ref{bubble-vE}) we can replace
$\omega_p^{2}(\xi)$ by $\omega_p^2(\xi =0)$. The validity of this
approximation for A15 compounds was discussed in
Ref.~\cite{Mitrovic1985PRB} and here we will assume that it is also
valid for HTSC. In this respect, the first assumption can be
considered as a statement that $\Sigma(\omega,\xi)$ is approximated
by $\Sigma(\omega,\xi=0)$ \cite{Allen2004}. After these two
assumptions are made we can integrate over $\xi$ and finally arrive
at the following representation for the optical conductivity (see
e.g. Refs.~\cite{Shulga1991PC,Lee1989PC,Kaufmann1998JSC,Allen2004})
\begin{equation}
\label{sigma-final} \sigma(\omega) = \frac{\omega_p^2}{4 \pi}
\frac{i}{\omega} \int_{-\infty}^{\infty} d \epsilon
\left[n_F(\epsilon)- n_F(\epsilon+\omega) \right]
\frac{1}{\omega+i/\tau_{\mathrm{imp}}+
\Sigma^{\ast}(\epsilon)-\Sigma(\epsilon+\omega)},
\end{equation}
where $\Sigma(\epsilon)$ is the retarded self-energy on the real
axes and $\Sigma^{\ast}(\epsilon)$ its complex conjugate. In
Eq.~(\ref{sigma-final}) we also included the electron-impurity
scattering rate, $1/\tau_{\mathrm{imp}}$, its frequency dependence
will be considered in Sec.~\ref{sec:tau-imp}. One can check that for $T=0$
Eq.~(\ref{sigma-final}) reduces to the expression for
$\sigma(\omega)$ written in
Refs.~\cite{Carbotte.book,Marsiglio1998PLA,Marsiglio1999JSC}.

To investigate the effect of electron-boson interaction on
$1/\tau(\omega)$ we must express $\mbox{Re}[1/\sigma(\omega)]$ in
terms of the self-energy $\Sigma$ in the simplest possible form.  It
can be anticipated if one substitutes Eq.~(\ref{sigma-final}) in
Eq.~(\ref{tau_via_sigma}) and rewrites it as follows:
\begin{equation}
\label{tau-anticipate} \frac{1}{\tau(\omega)} = \omega \mbox{Im}
\left\{ \frac{1}{\omega+i/\tau_{\mathrm{imp}}}
\int_{-\infty}^{\infty} d \epsilon
\left[n_F(\epsilon)-n_F(\epsilon+\omega)
 \right] \frac{1}{1- [\Sigma(\epsilon+\omega) -\Sigma^\ast(\epsilon)]/
(\omega+i/\tau_{\mathrm{imp}}) } \right\}^{-1}
\end{equation}
Now expanding the denominator of  Eq.~(\ref{tau-anticipate}), doing
the integration and then ``de-expanding'' the result (see
Refs.~\cite{Marsiglio1998PLA,Marsiglio1999JSC}) we obtain the
following approximate representation
\begin{equation}
\label{Sigma2tau} \frac{1}{\tau(\omega)} =
\frac{1}{\tau_{\mathrm{imp}}} - \frac{1}{\omega}
\int_{-\infty}^{\infty} d \epsilon \left[
n_F(\epsilon)-n_F(\epsilon+\omega)\right] \mbox{Im}
[\Sigma(\epsilon+\omega)- \Sigma^\ast(\epsilon)].
\end{equation}
In deriving Eq.~(\ref{Sigma2tau}) we used the assumption
$|\Sigma(\epsilon+\omega)-\Sigma^\ast(\epsilon)| \ll
|\omega+i/\tau_{\mathrm{imp}}|$ to expand and then assumed that
$$
\int_{-\infty}^{\infty} d \epsilon \left[
n_F(\epsilon)-n_F(\epsilon+\omega)\right]
|\Sigma(\epsilon+\omega)- \Sigma^\ast(\epsilon)| \ll \omega
|\omega+i/\tau_{\mathrm{imp}}|
$$
to ``de-expand''. Based on these inequalities one would expect that
all results that follow from  Eq.~(\ref{Sigma2tau})  are valid only
for large $\omega$. Nevertheless, the numerical comparison of the
results obtained by the direct substitution of the self-energy
(\ref{Sigma-Shulga}) in Eq.~(\ref{sigma-final}) with
$1/\tau(\omega)$ computed using Eq.~(\ref{tau-Shulga}) in
Ref.~\cite{Shulga1991PC} shows that the last equation is valid for a
much wider range of the frequency $\omega$. This is in spite of the
fact that the derivation of Eq.~(\ref{tau-Shulga}) is based on the
approximate Eq.~(\ref{Sigma2tau}). Finally we remind that since we
did not include vertex corrections, $1/\tau(\omega)$ is expressed in
terms of the usual self-energies instead of transport
``self-energies'' discussed in Ref.~\cite{Kaufmann1998JSC}.
Accordingly in Sec.~\ref{sec:tau-new} the optical scattering rated
$1/\tau(\omega)$ will be expressed in terms of the  tunneling
$\alpha^2 F$ instead of the transport $\alpha_{\mathrm{tr}}^2F$
considered by Allen in Ref.~\cite{Allen1971PRB}.

\section{Self-energy for $\tilde{N}(\xi)\ne \mbox{const}$}
\label{sec:self-energy}

Now we consider the influence of a nonconstant DOS
on the usual relationship between the self-energy
$\Sigma(\omega)$ and the electron-boson interaction function
$\alpha^2F$. We begin with the well-known expression (see e.g.
Refs.~\cite{Allen.review,Mitovic.thesis})
\begin{equation}
\label{Sigma-general}
\Sigma(i \omega_n) = T \sum_{m=-\infty}^{\infty}
\int_{-\infty}^{\infty} d \xi \frac{N(\xi)}{N(0)}
\int_{0}^{\infty} d \Omega \alpha^2 F(\Omega)
\frac{2\Omega}{\Omega_m^2+\Omega^2} G(i \omega_n-i \Omega_m,\xi)
\end{equation}
written for the case when $\alpha^2F(\Omega)$
does not depend on the electron energy. In Eq.~(\ref{Sigma-general})
\begin{equation}
\label{G_xi}
G(i\omega_n,\xi) \equiv \frac{1}{N(\xi)} \int \frac{d^2 k}{(2 \pi)^2}
\delta(\xi - \xi(\mathbf{k})) G(i \omega_n,\mathbf{k}),
\end{equation}
the DOS $N(\xi)$ is defined by Eq.~(\ref{DOS-band})
and its energy dependence usually is also neglected. Our goal is,
however, to retain $N(\xi)$ and consider the influence of $N(\xi)
\ne \mbox{const}$ on $\Sigma(\omega)$ and, accordingly, on
$\sigma(\omega)$. Again using the spectral representation for the
Green's function $G(i \omega_n,\mathbf{k})$, one can easily sum
over Matsubara frequencies in Eq.~(\ref{Sigma-general}) and obtain
\begin{equation}
\label{Sigma-summed} \Sigma(i \omega_n) = \int_{-\infty}^{\infty} d
\xi \frac{N(\xi)}{N(0)} \int_{-\infty}^{\infty} d \omega^\prime
\int_{0}^{\infty}d\Omega \alpha^2 F(\Omega) \left(-\frac{1}{\pi}
\mbox{Im} G_R(\omega+i0,\xi)\right) I(i
\omega_n,\Omega,\omega^\prime),
\end{equation}
where \cite{Allen.review}
\begin{equation}
I(i \omega_n,\Omega,\omega^\prime) =
\frac{n_B(\Omega)+1-n_F(\omega^\prime)}
{i\omega_n - \Omega - \omega^\prime} +
\frac{n_B(\Omega)+n_F(\omega^\prime)}{i\omega_n+\Omega -\omega^\prime}
\end{equation}
with the Bose distribution $n_B(\Omega)=1/(e^{\beta \Omega} -1)$.

Let us now consider the quantity
\begin{equation}
\tilde{N}(\omega) \equiv -\frac{1}{\pi} \int_{-\infty}^{\infty}
d \xi N(\xi) \mbox{Im} G_{R} (\omega+i0,\xi)
\end{equation}
which enters Eq.~(\ref{Sigma-summed}). Using Eq.~(\ref{G_xi}) and
the definition of the spectral function (\ref{SF}) one can easily
check that
\begin{equation}
\tilde{N}(\omega) = \int \frac{d^2 k}{(2\pi)^2} A(\omega,\mathbf{k}),
\end{equation}
viz. this quantity represents the fully dressed quasiparticle DOS
which could contain a pseudogap that has its origin in correlation
effects. Finally making an analytical continuation $i \omega_n \to
\omega+i0$ and taking the imaginary part of $\Sigma(\omega)$ we
obtain
\begin{equation}\label{ImSigma-general}
\mbox{Im} \Sigma(\omega) = - \pi \int_{0}^\infty d \Omega
\alpha^2F(\Omega) \left\{
\frac{\tilde{N}(\omega-\Omega)}{N(0)}[n_B(\Omega)+1-n_F(\omega-\Omega)]
+
\frac{\tilde{N}(\omega+\Omega)}{N(0)}[n_B(\Omega)+n_F(\omega+\Omega)]
\right\}.
\end{equation}
It is easy to see that for $\tilde{N}(\omega) = N(0) = \mbox{const}$
the previous equation reduces to a more familiar expression
\cite{Allen.review,Shulga1991PC}
\begin{equation}
\label{Sigma-Shulga} \mbox{Im}\Sigma(\omega)=-\frac{\pi}{2}
\int_{0}^\infty d \Omega \alpha^2F(\Omega) \left[ 2 \coth
\frac{\Omega}{2T}-\tanh \frac{\omega+\Omega}{2T}+\tanh
\frac{\omega-\Omega}{2T}\right].
\end{equation}

\section{Optical scattering rate: boson contribution}
\label{sec:tau-new}

Substituting the self-energy (\ref{ImSigma-general}) in the
expression (\ref{Sigma2tau}) and doing simple replacements of the
variables, we arrive at the main result of the present paper,
\begin{equation}
\label{tau-mine} \frac{1}{\tau(\omega)} = \frac{\pi}{\omega}
\int_{0}^{\infty} d \Omega \alpha^2 F(\Omega)
\int_{-\infty}^{\infty} d \epsilon \left[
\frac{\tilde{N}(\epsilon-\Omega)}{N(0)} +
\frac{\tilde{N}(-\epsilon+\Omega)}{N(0)} \right] [n_B(\Omega) +
n_F(\Omega-\epsilon)] [n_F(\epsilon-\omega) - n_F(\epsilon+\omega)],
\end{equation}
which establishes a link between the optical scattering rate
$1/\tau(\omega)$, the electron-boson interaction function $\alpha^2
F(\Omega)$ and the quasiparticle DOS $\tilde{N}(\omega)$. It is
important to stress that in contrast to $N(\omega)$, the
quasiparticle DOS $\tilde{N}(\omega)$ which enters
Eqs.~(\ref{ImSigma-general}) and (\ref{tau-mine}) is directly
related to the  spectral function $A(\omega,\mathbf{k})$ measured by
ARPES experiments \cite{Timusk1999RPP}. Note that in
Eq.~(\ref{tau-mine}) material parameters enter only as the
electron-boson spectral density and the fully dressed electron DOS.
Different mechanisms leading to the same $\tilde{N}(\omega)$ are
differentiated in the optical scattering rate only through the size
and shape of $\alpha^2 F(\Omega)$.

To understand better the rather lengthy Eq.~(\ref{tau-mine}) we
consider limiting cases where one can establish a link with
already known results.\\
\noindent (i) For $T=0$ the Bose distribution in
Eq.~(\ref{tau-mine}) drops out and  it  reduces to
Eq.~(\ref{tau-Mitrovic}) with the band DOS $N(\omega)$ replaced by
the quasiparticle DOS $\tilde{N}(\omega)$. Note that it was pointed
out in Ref.~\cite{Mitrovic1985PRB} that $N(\omega)$ in
Eq.~(\ref{tau-Mitrovic}) should be interpreted as the quasiparticle
DOS, but the golden rule approach of Allen \cite{Allen1971PRB} used
in  Ref.~\cite{Mitrovic1985PRB}  does not allow to establish this
fact directly. Comparing Eqs.~(\ref{tau-Mitrovic}) and
(\ref{tau-mine}) one can see that finite temperature brings an
essential  element, the Bose distribution $n_B(\Omega)$.

\noindent (ii) When $\tilde{N}(\omega) = \mbox{const}$
it becomes possible to integrate over $\epsilon$ in (\ref{tau-mine}).
Indeed using the integral
\begin{equation}
\int_{-\infty}^{\infty} d z n_F(z+a) n_F(-z-b) = (a-b)n_B(a-b)
\end{equation}
we arrive at Eq.~(\ref{tau-Shulga}) obtained by Shulga
et al. in Ref.~\cite{Shulga1991PC}.
Obviously,  Eq.~(\ref{tau-Shulga}) also follows directly from
Eqs.~(\ref{Sigma-Shulga}) and (\ref{Sigma2tau}).

\noindent (iii)  At temperatures much higher than the boson
spectrum upper-energy cutoff, $T\gg \Omega_c$, expression
(\ref{tau-mine}) reduces to
\begin{equation}
\label{tauN-highT} \lim_{T/\Omega_c \to
\infty}\frac{1}{\tau(\omega)} = \frac{\pi T}{\omega}
\int_{0}^{\infty} d\Omega \frac{\alpha^2 F(\Omega)}{\Omega}
\int_{-\infty}^{\infty} d \epsilon
\left[\frac{\tilde{N}(\epsilon-\Omega)}{N(0)}+
\frac{\tilde{N}(-\epsilon+\Omega)}{N(0)}\right]
[n_F(\epsilon-\omega)-n_F(\epsilon+\omega)].
\end{equation}
When $\tilde{N}(\epsilon) = \mbox{const}$ the last equation
can be further simplified \cite{Puchkov1996JPCM}
\begin{equation}
\label{tau-highT}
\lim_{T/\Omega_c \to \infty}\frac{1}{\tau(0)} =4 \pi T
\int_{0}^{\infty} d\Omega
\frac{\alpha^2 F(\Omega)}{\Omega} .
\end{equation}
In the case when the electron-phonon interaction is the origin of
$\alpha^2F(\Omega)$, Eq.~(\ref{tau-highT}) reflects the familiar
result that the high-temperature electron-phonon contribution to a
dc resistivity is linear in temperature. This no longer strictly
holds for Eq.~(\ref{tauN-highT}) where there is an additional $T$
dependence in the integral over $\epsilon$.

\section{Optical scattering rate: contribution of impurities}
\label{sec:tau-imp}

While Eq.~(\ref{tau-mine}) represents the electron-boson
contribution to the optical scattering rate, the total scattering
rate
\begin{equation}
\frac{1}{\tau_{\mathrm{tot}} (\omega)} =\frac{1}{\tau(\omega)}
+ \frac{1}{\tau_{\mathrm{imp}}(\omega)}
\end{equation}
consists of two parts, viz. the above-mentioned bosonic
$1/\tau(\omega)$ and that caused by impurities
$1/\tau_{\mathrm{imp}}(\omega)$.

There is a simple way of obtaining $1/\tau_{\mathrm{imp}}(\omega)$
by using the following expression for
\begin{equation}
\label{alpha-imp}
\alpha^2 F_{\mathrm{imp}}(\Omega) = \frac{1}{2 \tau_{\mathrm{imp}}}
\frac{\Omega \delta(\Omega)}{\pi T}
\end{equation}
and assuming that the integration of $\delta(\Omega)$ over $\Omega$
from 0 to $\infty$ gives $1/2$. $1/\tau_{\mathrm{imp}}$  in
Eq.~(\ref{alpha-imp}) is the normal state impurity scattering rate
and it is implied that the limit $\Omega \to 0$ in $\alpha^2
F_{\mathrm{imp}}(\Omega)$  must be taken before doing the limit $T
\to 0$. Substituting Eq.~(\ref{alpha-imp}) in Eq.~(\ref{tau-mine})
we arrive at the expression
\begin{equation}
\label{tau-imp} \frac{1}{\tau_{\mathrm{imp}}(\omega)} =
\frac{1}{\tau_{\mathrm{imp}}} \frac{1}{4 \omega}
\int_{-\infty}^{\infty} d \epsilon [n_F(\epsilon-\omega)-
n_F(\epsilon+\omega)] \left[\frac{\tilde{N}(\epsilon)}{N(0)}+
\frac{\tilde{N}(-\epsilon)}{N(0)} \right].
\end{equation}
Energy dependence in $\tilde{N}(\epsilon)$ implies an energy and
temperature dependence in the impurity scattering rate. For $T=0$
Eq.~(\ref{tau-imp}) reduces to the result of
Ref.~\cite{Mitrovic1985PRB}
\begin{equation}
\label{tau_imp-Mitrovic} \frac{1}{\tau_{\mathrm{imp}}(\omega)} =
\frac{1}{\tau_{\mathrm{imp}}} \frac{1}{\omega} \int_{0}^{\omega} d
\epsilon \frac{1}{2}\left[\frac{\tilde{N}(\epsilon)}{N(0)}+
\frac{\tilde{N}(-\epsilon)}{N(0)} \right].
\end{equation}
Note that due to the above-mentioned noncommutativity of the limits
$\Omega\to 0$ and $T \to 0$ in Eq.~(\ref{alpha-imp}), the last
expression cannot be obtained simply by substituting
Eq.~(\ref{alpha-imp}) in Eq.~(\ref{tau-Mitrovic}). It is very easy
to see that for $\tilde{N}(\epsilon)= \mathrm{const}$,
Eq.~(\ref{tau_imp-Mitrovic}) as well as Eq.~(\ref{tau-imp}) reduce
to the trivial result $\tau_{\mathrm{imp}}(\omega) =
\tau_{\mathrm{imp}}$. For a more extensive discussion of the
impurity scattering problem when the DOS is energy dependent the
reader is referred to Ref.~\cite{Li2002PRB}. In the fits to data
made in Ref.~\cite{Hwang2005} residual scattering is small and not
important. In that case $1/\tau_{\mathrm{imp}}$ is estimated to be a
few meV while the inelastic scattering rate rises to a value larger
than 600 meV in the frequency range of interest in the fit. Under
such circumstances any modulation of $1/\tau_{\mathrm{imp}}(\omega)$
brought about by energy dependence in $\tilde{N}(\epsilon)$ is of no
consequence when fitting to the inelastic scattering part. This
would not be so if impurity scattering became large and of the order
of the inelastic scattering in the important frequency range.

\section{Discussion}
\label{sec:discussion}

\begin{figure}[h]
\centering{
\includegraphics[width=8cm]{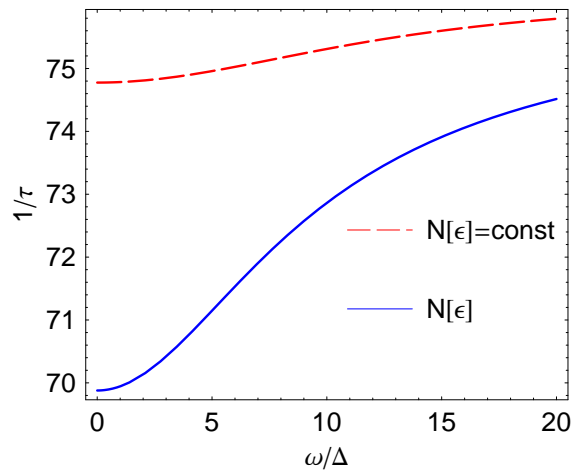}}
\caption{(Color online) The dependence $1/\tau(\omega)$ (in
arbitrary units) obtained from Eqs.~(\ref{tau-Shulga}) for
$\tilde{N}(\epsilon) = \mbox{const}$ and (\ref{tau-mine}) with
$\tilde{N}(\epsilon)$ given by Eq.~(\ref{DOS-PG}). We take the
Einstein model for $\alpha^2 F(\Omega) = \alpha^2
\delta(\Omega-\Omega_E)$ with $\Omega_E = 2 \Delta$ and $T=2.5
\Delta$.} \label{fig:1}
\end{figure}

To illustrate the effect of the opening of the pseudogap on
$1/\tau(\omega)$ in Fig.~\ref{fig:1} we plot $1/\tau(\omega)$
computed with and without pseudogap, but do not consider the
contribution from impurities. The case without pseudogap which
corresponds to $\tilde{N}(\epsilon) = N(0) = \mbox{const}$ was
considered using Eq.~(\ref{tau-Shulga}), while to model a pseudogap
we took \cite{Hwang2005}
\begin{equation}\label{DOS-PG}
\frac{\tilde{N}(\epsilon)}{N(0)} = \left[\frac{\tilde{N}(0)}{N(0)} +
\left(1-  \frac{\tilde{N}(0)}{N(0)}\right)
\frac{\epsilon^2}{\Delta^2}\right] \theta(\Delta-|\epsilon|) +
\theta(|\epsilon|-\Delta)
\end{equation}
and used  Eq.~(\ref{tau-mine}). It is easy to see that the main
effect of the opening of the pseudogap [e.g., $\tilde{N}(0) < N(0)$]
is to reduce $1/\tau(\omega)$. This result implies that when there
is a pseudogap one cannot use Eq.~(\ref{tau-Allen}) to estimate
$\alpha^2F(\Omega)$, because the strength of the peaks of
$\alpha^2F(\Omega)$ would be underestimated. More importantly,  the
position of these peaks would be shifted  depending on our
assumptions about the absence or presence of the pseudogap.

Indeed by applying Eq.~(\ref{tau-mine}) in Ref.~\cite{Hwang2005} to
the analysis of the experimental data it was demonstrated that there
is agreement between the sharp bosonic resonance observed in the
infrared scattering rate and the properties of the $(\pi,\pi)$
spin-flip neutron mode. Using the above-mentioned property that the
decrease of $N(\epsilon)$ due to the opening of the pseudogap
effectively reduces $1/\tau(\omega)$, one can choose the position of
the bosonic resonance in $\alpha^2F(\Omega)$ at $\Omega = 248
\mbox{cm}^{-1}$ and thus makes it agree with the frequency of the 31
meV neutron mode, measured by Stock {\em et al.}
\cite{Stock2004PRB}. The use of Eq.~(\ref{tau-mine}) is crucial for
this agreement and the conventional relationship (\ref{tau-Allen})
valid for $T=0$ and $N(\epsilon) = \mbox{const}$ would produce a
peak in $\alpha^2F(\Omega)$ at $\Omega = 350 \mbox{cm}^{-1}$. The
possibility of achieving a reconciliation between the experimental
results obtained from neutron scattering and optical conductivity is
quite important in developing a coherent theoretical description of
cuprate superconductors and shows the usefulness of
Eq.~(\ref{tau-mine}).

We note  that since this kind of analysis involves modeling the form
of the pseudogap and its temperature dependence, the final results
for $\alpha^2F(\Omega)$ are definitely not unique. However, one does
not expect any qualitative change to the model for
$\alpha^2F(\Omega)$ obtained in Ref.~\cite{Hwang2005} which includes
a very wide temperature independent background and a sharp
temperature-dependent peak. As a given sample is studied using
several experimental techniques and the quality of the data
improves, the fit should  become more unique. Also it would be
interesting to repeat the analysis made in
Ref.~\cite{Dordevic2005PRB}, but now based on  Eq.~(\ref{tau-mine})
derived in this paper, rather than on Eq.~(\ref{tau-Shulga}) which
is valid only for a constant density of states.

Finally we mention that in Ref.~\cite{Dordevic2005PRB} data were
analyzed not only in the normal or pseudogap state at finite $T$,
but also in the superconducting state. This latter analysis is based
on a relationship between $1/\tau(\omega)$ and $\alpha^2F(\Omega)$
derived in Ref.~\cite{Allen1971PRB} for $s$-wave superconductor at
$T=0$. In relation to the results obtained in the present paper it
is worthwhile to ask the following question.  Is it possible to
distinguish the decrease of $1/\tau(\omega)$ caused by the pseudogap
and by the superconducting $s$- or even $d$-wave gap? To address
this question in detail there is a need to generalize the
corresponding expression from Ref.~\cite{Allen1971PRB} to $d$-wave
symmetry of the superconducting gap and $T \neq 0$. Such
considerations go beyond the scope of the present work.

\section{Acknowledgments}
The authors gratefully acknowledge J.~Hwang, A.~Knigavko,
F.~Marsiglio, and T.~Timusk for helpful discussions. This work was
supported by the Natural Science and Engineering Council of Canada
(NSERC) and by the Canadian Institute for Advanced Research (CIAR).

\end{document}